\documentclass[prd,twocolumn]{revtex4}%
\usepackage{amsfonts}
\usepackage{amsmath}
\usepackage{amssymb}
\usepackage{graphicx}%
\begin{document}
\title{Influence of gravitational waves on quantum multibody states}
\author{Jiatong Yan}
\author{Baocheng Zhang}
\email{zhangbaocheng@cug.edu.cn}
\affiliation{School of Mathematics and Physics, China University of Geosciences, Wuhan
430074, China}
\keywords{Gravitational wave, Twin-Fock state, Unruh-DeWitt detector }
\pacs{04.30.¨Cw, 03.75.Gg, 04.62.+v}

\begin{abstract}
Based on the freely-falling Unruh-Dewitt model, we study the influence of
gravitational waves on the quantum multibody states, i.e. the twin-Fock (TF)
state and the mixture of Dicke states. The amount of entanglement of quantum
many-body states decreases first and then increases with increasing frequency
of gravitational waves. In particular, for some fixed frequencies of
gravitational waves, entanglement will increase with the increasing amplitude
of gravitational waves, which is different from the usual thought of
gravity-induced decoherence and could provide a novel understanding for the
quantum property of gravitational waves.

\end{abstract}
\maketitle

\section{Introduction}

Quantum information is a fascinating subject which has the capacity to
revolutionize our understanding of the Universe, and it has been applied as a
tool to understand some relativistic phenomena in a variety of different
settings such as the acceleration and black holes \cite{pt08,chm08} (known as
the Unruh and Hawking effects).

Quantum entanglement, as the most interesting feature of quantum information,
has been used as a method to enhance the sensitivity of gravitational wave
detectors. References \cite{mmc17,sss20} studied the feasibility of
eliminating the need of the filter cavity by harvesting mutual quantum
correlations and discussed the difference in the way each beam propagates in
the interferometer. Reference \cite{edk18} proposed a new implementation using
a quantum speedmeter measurement scheme for gravitational wave detection based
on quantum entanglement. Apart from these, some papers studied in principle
about quantum properties affected by gravitational waves, including quantum
imprints \cite{gkt21}, quantum time dilation \cite{pdd22}, entanglement
harvesting \cite{xsa20}, excitation/deexcitation of a single atom
\cite{tp22,hp22}, and so on. The influence of gravitational field on quantum
entanglement was also studied in \cite{yzy22}. But most of these studies
concentrated on the two-body entanglement. In this paper, we will study the
influence of gravitational waves on quantum many-body states and discuss the
feasibility of experimental detection for gravitational waves.

A recent protocol of using Bose-Einstein condenstate (BEC) as a gravitational
wave detector has been proposed in \cite{sbf14,ram19,ram22}. By adjusting the
squeezing parameter of a single-mode BEC, the size of the BEC and the
observation time appropriately, the information about gravitational wave could
be distilled by calculating the Fisher information, but the sensitivity of
such a gravitational wave detector is far from enough in the current
experimental and technologic conditions. In this paper, we will employ the
recent experimentally realized quantum many-body states called as twin-Fock
(TF) state for the BEC to investigate the change of entanglement caused by
gravitational waves based on the Unruh-DeWitt (UDW) model \cite{wgu76,udm},
which is different from the earlier suggestions \cite{sbf14,ram19,ram22} in
the fundamental interaction. We will also explore the feasibility of detecting
gravitational wave using TF states.

This paper is organized as follows. In Sec. 2, we review the model of
Unruh-Dewitt detector and calculate the transition rate in monochromatic
nonpolarized gravitational waves background. This is followed in Sec. 3 by the
introduction of a certain kind of many-body state and the evolution of TF
states affected by gravitational waves, and the entanglement-related
spin-squeezing parameter will be calculated to show the change of quantum
many-body state. In Sec. 4, we discuss the experimental feasibility of using
TF state to detect gravitational waves. Finally, we summarize and give the
conclusion in Sec. 5.

\section{Unruh-Dewitt detector model}

In this section, we discuss a single UDW detector accelerated in the field of
gravitational waves. Start by considering that the gravitational waves
propagates through a flat spacetime with a perturbed metric given as,
\begin{equation}
g_{\mu\nu}=\eta_{\mu\nu}+h_{\mu\nu}%
\end{equation}
where $h_{\mu\nu}$ is the linear-order perturbation of the metric tensor
$g_{\mu\nu}$ around the flat Minkowski space characterized by $\eta_{\mu\nu
}=diag(-1,1,1,1)$. In the traceless-transverse gauge \cite{mtw73}, the
perturbation can be expressed as
\begin{equation}
h(u)=h_{+}\epsilon_{ij}^{+}cos(\omega_{g}u+\psi_{+})+h_{\times}\epsilon
_{ij}^{\times}cos(\omega_{g}u+\psi_{\times}), \label{sgw}%
\end{equation}
where $u=t-z$ is a light cone coordinate, $\omega_{g}$ is the frequency of the
gravitational wave, and $\psi_{+}$ ($\psi_{\times}$) is the initial phase for
the $+$ ($\times$) polarization component of the gravitational wave. $h_{+}$
and $h_{\times}$ describe two dimensionless amplitudes of a gravitational wave
for its two independent polarization components, respectively. $\epsilon
_{ij}^{+}$ and $\epsilon_{ij}^{\times}$ are the so-called unit polarization
tensors for the corresponding polarization components. In the paper, for
simplicity, we take the natural units in which $c=\hbar=1$. When we discuss
the effects of gravitational waves on specific quantum multibody states for
the experimental feasibility, the standard units will be restored.

Now, we consider that a freely falling UDW detector couples to the massless
scalar field fluctuating in the presence of gravitational waves. The
interaction Hamiltonian can be expressed as \cite{chm08}
\begin{equation}
H_{int}=-\lambda m(t)\phi(x),
\end{equation}
where $m(t)$ is the monopole moment of the detector and $\lambda$ a coupling
constant. At the first order of perturbation theory, the transition amplitude
from the ground state, $|E_{0}\rangle\otimes|\Omega\rangle$ (where
$|E_{0}\rangle$) denotes the ground state of the detector with energy $E_{0}$
and $|\Omega\rangle$ denotes the ground state of the scalar field) to a state
$|E\rangle\otimes|\Psi\rangle$, (where $|E\rangle$ is an excited state of the
detector with energy $E\geq E_{0}$), is given by
\begin{equation}
A(E_{0}\rightarrow E)=i\lambda\int_{-\infty}^{\infty}\langle E|\hat{m}%
(\tau)|E_{0}\rangle\langle\Psi|\hat{\phi}(x(\tau)|\Omega\rangle d\tau,\
\end{equation}
where $\tau$ is the detector's proper time and $x^{\mu}(\tau)$ is the
detector's world line. The probability $P$ that the detector transits from
$E_{0}$ to $E$ is obtained by squaring the transition amplitude, and summing
over all intermediate (excited) states of the field $|\Psi\rangle$, resulting
in
\begin{equation}
P=\lambda^{2}\sum_{E}|\langle E|\hat{m}(\tau)|E_{0}\rangle|^{2}F(\Delta E),
\label{tp}%
\end{equation}
where $\hat{m}(\tau)=e^{-i\hat{H}_{0}\tau}\hat{m}(0)e^{i\hat{H}_{0}\tau}$,
represents the evolution of the monopole for the atom with the Hamiltonian
$\hat{H}_{0}$ ($\hat{H}_{0}|E_{0}\rangle=E_{0}|E_{0}\rangle$, $\hat{H}%
_{0}|E\rangle=E|E\rangle$) and $F(\Delta E)$ denotes the response function of
the detector given by
\begin{equation}
F(\Delta E)=\int_{-\infty}^{\infty}d\tau\int_{-\infty}^{\infty}d\tau^{\prime
}e^{-\frac{i}{\hbar}\Delta E(\tau-\tau^{\prime})}W(x^{\mu}(\tau);x^{\nu}%
(\tau^{\prime})),
\end{equation}
where $\Delta E=E-E_{0}$ and $W(x^{\mu}(\tau);x^{\nu}(\tau^{\prime}))$ is the
positive frequency Wightman function evaluated along the geodesics of the
detector. For the scalar field $\hat{\phi}$ that satisfies the Klein-Gordon
equation, the positive Wightman functions are defined as the following
two-point functions \cite{bd84,st86}, $W(x^{\mu}(\tau);x^{\nu}(\tau^{\prime
}))=\langle\Omega|\hat{\phi}(x)\hat{\phi}(x^{\prime})|\Omega\rangle$.

Employing variable substitutions $\Delta\tau=\tau-\tau^{\prime}$ and
$T=(\tau+\tau^{\prime})/2$ ($\Delta\tau$ is the relative time and $T$ is the
average time),
\begin{equation}
F(\Delta E)=\int_{-\infty}^{\infty}dT\int_{-\infty}^{\infty}d\Delta\tau
e^{-\frac{i}{\hbar}\Delta E\Delta\tau}W(x^{\mu};x^{\nu}), \label{wfe}%
\end{equation}
where $W(x^{\mu};x^{\nu})\equiv W(x^{\mu}(T+\Delta\tau/2);x^{\nu}(T-\Delta
\tau/2))$, such that one can define the rate $R$ per unit time for the
response function of the detector,
\begin{align}
R(T,\Delta E)  &  =\lim_{\Delta T\rightarrow0}[\frac{\Delta F(\Delta
E)}{\Delta T}]\nonumber\\
&  =\int_{-\infty}^{\infty}d\Delta\tau e^{-\frac{i}{\hbar}\Delta E\Delta\tau
}W(x^{\mu};x^{\nu}). \label{rpt}%
\end{align}

In order to calculate the rate $R$, the geodesics of the free-falling UDW
detector in the background of gravitational waves has to be given. Here we
write the line element as
\begin{equation}
ds^{2}=-dudv+g_{ij}(u)dx^{i}dx^{j}, \label{len}%
\end{equation}
where $u=t-z$, $v=t+z$ are the light cone coordinates, and $g_{ij}(u)$ is a
$2\times2$ dimensional matrix. Useful Killing vectors are $K_{\nu}%
=\partial_{\nu}$ and $K_{i}=\partial_{i}(i=1,2)$, from which one obtains the
corresponding conserved momenta, $P_{v}=-(K_{v})_{\mu}\frac{dx^{\mu}}{d\tau
}=\frac{1}{2}\frac{du}{d\tau}$,$\quad P_{i}=(K_{i})_{j}\frac{dx^{j}}{d\tau
}=g_{ij}(u)\frac{dx^{j}}{d\tau}$,$\quad(i,j=1,2)$, where we chose the geodesic
time $\lambda=\tau$ to be the proper time $\tau$, defined by $d\tau
^{2}=-ds^{2}$. With these conserved momenta and the line element (\ref{len}),
it is not hard to obtain the equations of geodesics,
\begin{align}
u(\tau)  &  =u_{0}+2P_{v}\tau,\quad(u_{0}\equiv u(0)),\label{geou}\\
v(\tau)  &  =v_{0}+\frac{1}{2P_{v}}(\tau+P_{i}(x^{i}(u)-x_{0}^{i}%
)),\label{geov}\\
x^{i}(\tau)  &  =x_{0}^{i}+\frac{1}{2P_{v}}\int_{u_{0}}^{u}d\overline{u}%
g^{ij}P_{j},\quad(x_{0}^{i}\equiv x^{i}(0)). \label{geox}%
\end{align}

With the equations of geodesics, we can continue to calculate the Wightman
function. The Wightman function in the gravitational wave background is given
as \cite{hp22}
\begin{equation}
W=\frac{m^{2}}{(2\pi)^{2}[\gamma(u)\gamma(u^{\prime})]^{\frac{1}{4}}%
\sqrt{\Gamma(u;u^{\prime})}}\frac{K_{1}(m\sqrt{\Delta\overline{x}^{2}}%
)}{m\sqrt{\Delta\overline{x}^{2}}},
\end{equation}
where $K_{\nu}(z)$ denotes Bessel's function of the second kind, and
$\Delta\overline{x}_{(\pm)}^{2}(x;x^{\prime})$ are the deformed distance
functions, which in light cone coordinates can be written as \cite{hp22},
\begin{equation}
\Delta\overline{x}^{2}(x;x^{\prime})=\left(
\begin{array}
[c]{cc}%
\Delta x & \Delta y
\end{array}
\right)  g(u;u^{\prime})\left(
\begin{array}
[c]{c}%
\Delta x\\
\Delta y
\end{array}
\right)  -(\Delta u-i\epsilon)(\Delta v-i\epsilon). \label{ddf}%
\end{equation}
This global Lorentz violation is mediated by the (symmetric) deformation
matrix.
\begin{equation}
g(u;u^{\prime})=\left(
\begin{array}
[c]{cc}%
g_{xx}(u;u^{\prime}) & g_{xy}(u;u^{\prime})\\
g_{xy}(u;u^{\prime}) & g_{yy}(u;u^{\prime})
\end{array}
\right)  ,
\end{equation}
which is the inverse of the corresponding momentum space deformation matrix
$\Gamma_{M}(u;u^{\prime})$, $g(u;u^{\prime})=\Gamma_{M}^{-1}(u;u^{\prime})$.

For monochromatic nonpolarized gravitational waves as in Eq. (\ref{sgw}) but
taking $h_{+}=h_{\times}$ and $\psi_{+}=0,\psi_{\times}=-\pi/2$ for
simplicity, one obtains
\begin{equation}
g(u;u^{\prime})=\frac{1}{\gamma\Gamma(u;u^{\prime})}\left(
\begin{array}
[c]{cc}%
1+hB_{1} & -hB_{2}\\
-hB_{2} & 1-hB_{1}%
\end{array}
\right)  , \label{dgw}%
\end{equation}%
\begin{equation}
\Gamma(u;u^{\prime})=det[\Gamma_{M}]=\frac{1}{\gamma^{2}}[1-h^{2}j_{0}%
^{2}(\frac{\omega_{g}\Delta u}{2})], \label{mgw}%
\end{equation}
where $j_{0}(z)=\sin(z)/z$ is the spherical Bessel function, $\gamma=1-h^{2}$
is time independent, $B_{1}=\frac{\sin{(\omega_{g}u)}-\sin{(\omega
_{g}u^{\prime})}}{\omega_{g}\Delta u}$, and $B_{2}=\frac{\cos{(\omega_{g}%
u)}-\cos{(\omega_{g}u^{\prime})}}{\omega_{g}\Delta u}$.

Now inserting the equations of geodesics (\ref{geou}), (\ref{geov}),
(\ref{geox}) into deformed distance functions (\ref{ddf}), and then inserting
Eqs. (\ref{ddf}), (\ref{dgw}) and (\ref{mgw}) into the expression (\ref{rpt})
of the rate, we get
\begin{equation}%
\begin{split}
&  R(\Delta E)=\frac{\hbar m^{2}\sqrt{\gamma}}{(2\pi)^{2}}\int_{-\infty
}^{\infty}\frac{d\Delta\tau}{[1-h^{2}j_{0}^{2}(P_{v}\omega_{g}\Delta
\tau)]^{1/2}}e^{-\frac{i}{\hbar}\Delta E\Delta\tau}\\
&  \times\frac{K_{1}({im(\Delta\tau-i\epsilon)})}{im(\Delta\tau-i\epsilon)},
\end{split}
\end{equation}
where we have made use of, $\Delta u=2P_{v}\Delta\tau$. The integrand in
Eq.(18) has two square-root cuts along the imaginary axis of complex
$\Delta\tau$, starting at the root of the equation, $j_{0}(P_{v}\omega
_{g}\Delta\tau)=\frac{\sinh(\theta)}{\theta}=\frac{1}{h}\gg1$, where
$\theta=-iP_{v}\omega_{g}\Delta\tau$. The root $\theta_{0}(h)$ can be
approximated by the solution of $\theta_{0}/(e^{\theta_{0}}-e^{-\theta_{0}%
})=h/2$. Equation (18) can be recast to
\begin{equation}
R(\Delta E)=\frac{\hbar m\sqrt{\gamma}}{2\pi^{2}}\int_{\theta_{0}}^{\infty
}\frac{d\theta}{[h^{2}\sinh^{2}{(\theta)}]^{1/2}}e^{-\frac{\Delta E}{\hbar
P_{v}\omega_{g}}\theta}K_{1}(\frac{m}{P_{v}\omega_{g}}\theta),
\end{equation}
which in the massless limit this reduces to
\begin{equation}
R(\Delta E)=\frac{\hbar\sqrt{\gamma}P_{v}\omega_{g}}{2\pi^{2}}\int_{\theta
_{0}}^{\infty}\frac{d\theta}{\theta\lbrack h^{2}\sinh^{2}(\theta)-\theta
^{2}]^{1/2}}e^{-\frac{\Delta E}{\hbar P_{v}\omega_{g}\theta}}.
\end{equation}
This is the numerical integration we need to perform. In this paper, we only
consider the massless case, and discuss the behaviors of the accelerated
quantum many-body states in the background of gravitational waves. Finally, we
want to stress that the calculation in this section is made for the initial
atomic state is the ground state, as seen in Eq. (\ref{tp}). Actually, the
calculation can be made for the initial atomic state is the excited, in which
the difference lies in the response function with $-\Delta E$ instead of
$\Delta E$ and all other calculation is similar. For distinguishing them, we
call $P_{+}$ and $R_{+}$ for case in which the initial state is the ground
state, and $P_{-}$ and $R_{-}$ for case in which the initial state is the
excited state.

\section{Quantum many-body states}

We choose the TF states as the many-body entangled quantum states and discuss
the influence of gravitational waves in this section. TF states are one kind
of Dicke states \cite{rhd54}. For a collection of $N$ identical (pseudo)
spin-1/2 particles, Dicke states can be expressed in Fock space as $|\frac
{N}{2}+m\rangle_{\uparrow}|\frac{N}{2}-m\rangle_{\downarrow}$ with $(\frac
{N}{2}+m)$ particles in spin-down modes for $m=-\frac{N}{2},-\frac{N}%
{2}+1,...,\frac{N}{2}$. In particular, $m=0$ represents just the TF state
where the number of the particles is the same for each one of the two-spin
states. On the other hand, Dicke states can be described by the common
eigenstate $|j,m\rangle$ of the collective spin operators $J^{2}$ and $J_{z}$,
with respective eigenvalues $j(j+1)$ and $m$. For the system consisted of $N$
two-level atoms we will consider, the state $|j=\frac{N}{2},m\rangle$
indicates the $(j+m)$ atoms are at the excited state $|e\rangle$, $(j-m)$
atoms are at the ground state $|g\rangle$.

The influence of gravitational waves on TF states will be investigated based
on such consideration that the single atom influenced by gravitational waves
will be calculated according to the Unruh-Dewitt detector and all atoms feel
the same gravitational waves without any other interaction among atoms except
their initial entanglement. Meanwhile, we neglect the distance between atoms
since it is much less than the relevant wavelengths of the field, with assures
that all atoms see the same field. When all atoms are influenced by
gravitational waves, the state of every atom is changed. Because of the
existence of the excitation probability and the deexcitation probability due
to the gravitational waves, the resulted atom's number at the excited states
$|e\rangle$ and the ground states $|g\rangle$ may not be equal. Thus, the
final state may deviate from the original state.

Considering the Unruh-Dewitt model for a single atom and the interaction
unitary operation $U=I-i\int d\tau H_{I}(\tau)+O(\lambda^{2})$ is expanded to
the first order. with the influence of gravitational waves on atoms in vacuum,
the evolution of the atoms can be written with the density operators as
\begin{equation}
\rho_{f}=U^{\dagger}\rho_{o}U,
\end{equation}
where the initial density operator consists of the product form of the density
operator for the atom and the density operator for the vacuum field, $\rho
_{v}=|0\rangle\langle0|$. Thus, within the first-order approximation and in
the interaction picture, the evolution of the atom could be described by
\begin{align}
Tr_{v}[U^{\dagger}(\rho_{a1}\otimes\rho_{v})U]  &  =|g\rangle\langle
g|+P_{+}|e\rangle\langle e|,\label{fe1}\\
Tr_{v}[U^{\dagger}(\rho_{a2}\otimes\rho_{v})U]  &  =|e\rangle\langle
e|+P_{-}|g\rangle\langle g|, \label{fe2}%
\end{align}
where $Tr_{v}$ represents the calculation of tracing out the field degrees of
freedom, and $P_{+}$ and $P_{-}$ are excitation and deexcitation
probabilities. The initial density operators for the atom are taken as
$\rho_{a1}=|g\rangle\langle g|$, $\rho_{a2}=|e\rangle\langle e|$. These
expression in Eqs. (\ref{fe1}) and (\ref{fe2}) provide the elementary forms of
the evolution for the general state of the atoms.

According to the Fermi's golden rule, the transition probabilities $P_{\pm}$
should be proportional to the transition rates $R_{\pm}$, and for a total
observation time of $t_{obs}$, it can be taken as $P_{\pm}=t_{obs}R_{\pm}$
\cite{pt08}. Now, we extend this to a system of $N$ atoms. We take the initial
many-body state as a standard TF state $|j,0\rangle$. When all atoms are
influenced by the same gravitational waves, the TF state $|j,0\rangle$
becomes
\begin{equation}
\rho_{t}=\sum_{m=-N/2}^{N/2}A_{m}^{2}|j,m\rangle\langle j,m|, \label{ntf}%
\end{equation}
where $A_{m}^{2}=[\sum_{k=0}^{N/2-|m|}C_{N/2}^{k}C_{N/2}^{k+|m|}(P_{+}%
P_{-})^{k}(\theta(m)(P_{+})^{m}+\theta(-m)(P_{-})^{|m|}$ in which the function
$\theta(x)=1$ when $x>0$ and $\theta(x)=0$ otherwise, and $C_{n}^{r}=\frac
{n!}{r!(n-r)!}$ denotes the combinatorial factor of choosing $r$ out of $n$.
The parameter $A_{0}^{2}$ represents the probability of remaining in the
original TF state, which includes those cases that if $l(0\leq\frac{N}{2})$
atoms are changed from the ground states to the excited states, there must be
other $l$ atoms which are changed from the excited states to the ground states
simultaneously. The second term appears due to the inequality between the
excitation probability $P_{+}$ and deexcitation probability $P_{-}$. The
parameter $A_{m}^{2}$ can be worked out by choosing the terms that in every
term either there are $m$ more excited states than ground states (that is the
case for $m>0$) or there are $m$ more ground states than excited states (that
is the case for $m<0)$. The crossed terms like $|a,m\rangle\langle
a,m^{\prime}|$ have been reduced when tracing out the field degrees of freedom.

We choose the spin-squeezing parameter to show the change of the original
quantum many-body state \cite{mwn11,sdz01,tkp09,gt09},
\begin{equation}
\xi_{E}^{2}=\frac{(N-1)(\Delta J_{z})^{2}+\langle J_{z}^{2}\rangle}{\langle
J^{2}\rangle-N/2},
\end{equation}
where the mean-spin direction was taken along the $z$ direction. If $\xi
_{E}^{2}<1$, the state is spin squeezed and entangled. The smaller the value
of $\xi_{E}^{2}$, the more the entanglement will be. In particular, in our
paper, $\xi_{E}^{2}=0$ represents the most spin-squeezed and entangled TF
state. Under the influence of gravitational waves, the multibody states will
deviate from the original state and the spin squeezing will be slightly
changed. With the evolved TF state (\ref{ntf}) influenced by the gravitational
waves, we can calculate $(\Delta J_{z})^{2}=\langle J_{z}^{2}\rangle-\langle
J_{z}\rangle^{2}$ and $\langle J\rangle=\frac{N}{2}(\frac{N}{2}+1)$ with
$\langle J_{z}\rangle=Tr(\rho_{t}J_{z})$ and $\langle J_{z}^{2}\rangle
=Tr(\rho_{t}J_{z}^{2})$.

The resulted spin-squeezing parameter is shown in Fig. 1 as a function of the
frequency of gravitational wave $f$ ($2\pi f=\omega_{g}$). It is seen from
this figure that under the influence of gravitational waves, the quantum
many-body state will deviate from its original state, but the spin-squeezing
parameter first increases and then decreases with increasing frequency. This
means that at the frequency where a peak appears in the curve of the
spin-squeezing parameter, the quantum many-body state feels the greatest
influence by the gravitational waves. \begin{figure}[ptb]
\centering
\includegraphics[width=1\columnwidth]{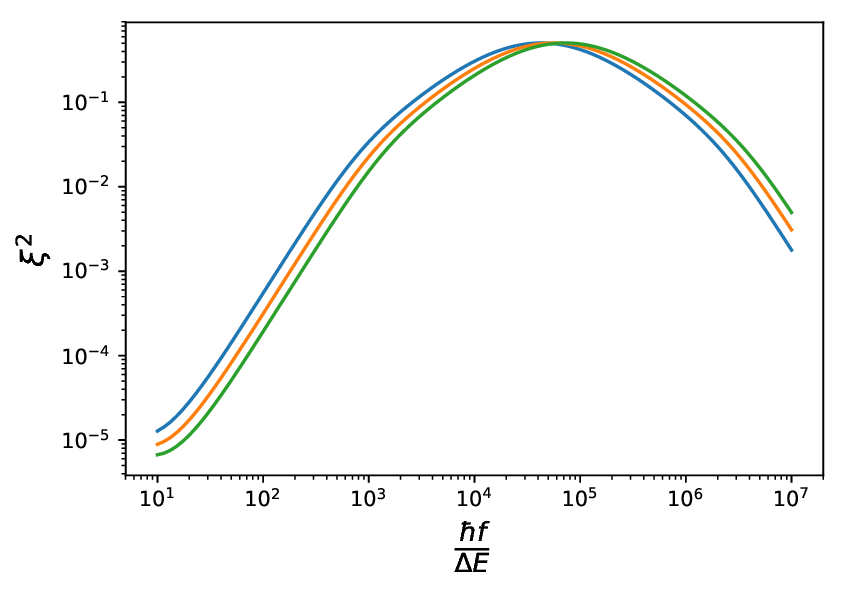}\caption{A plot of the phase
sensitivity of quantum multibody states influenced by gravitational waves for
different amplitude $h$. We set $P_{v}=1$, $t_{obs}=1$, and $N=50$. Blue,
orange and green lines stand for $h=10^{-18}$, $h=10^{-21}$ and $h=10^{-24}$
respectively.}%
\label{Fig1}%
\end{figure}

How about the change of $\xi^{2}$ with respect to amplitude $h$? We draw the
lines in Fig. 2 to show the change of $\xi^{2}$ with respect to $h$ for
different frequencies. It is clear from this figure that, for small frequency
gravitational wave, $\xi^{2}$ increases with $h$ growing, but for large
frequency, $\xi^{2}$ decreases with $h$ growing, which means that quantum
entanglement may increase with increasing amplitude of gravitational waves.
This phenomenon is counterintuitive but analogous to the celebrated anti-Unruh
phenomena \cite{bmm16,lzy18} where entanglement could also be amplified with
increasing acceleration. This is a novel phenomena which has not been found
before, and show that entanglement is increased by the increasing
gravitational magnitude, which is different from the view from the
gravitational decoherence \cite{bgu17}.

\begin{figure}[ptb]
\centering
\includegraphics[width=1\columnwidth]{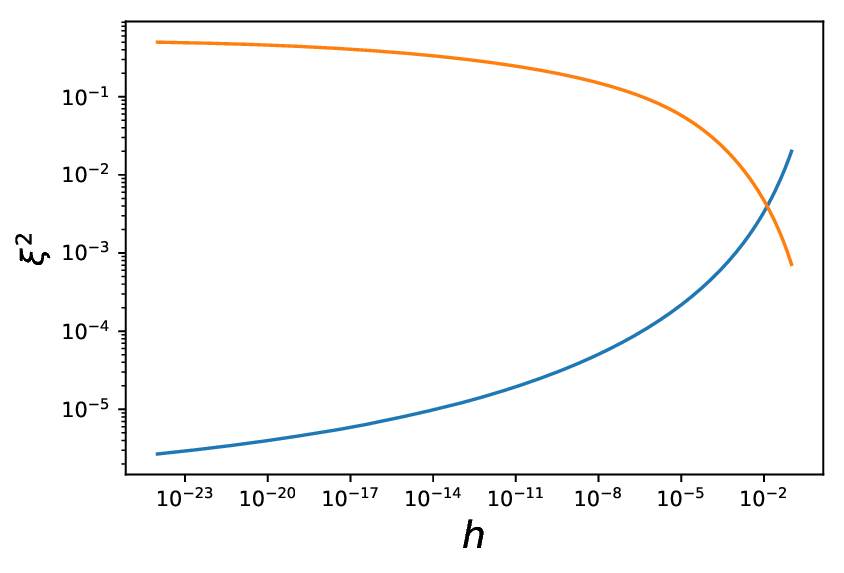}\caption{A plot of the
spin-squeezing parameter of quantum multibody states influenced by
gravitational waves as a function of the amplitude $h$ for different
frequencies. Blue and orange lines stand for $f = 1$ and $f = 10^{4}$,
respectively. $N = 20$ and other parameters are the same as Fig. 1.}%
\label{Fig2}%
\end{figure}

\section{Experimental possibility}

In order to check the size of the influence from the gravitational waves, we
consider the experiment of the Ramsey interferometer \cite{nr85,ymk86} with
the initial input state $\rho_{i}$, and the output state $\rho_{o}%
=U_{R}^{\dagger}\rho_{i}U_{R}$ where $U_{R}=exp(-i\theta J_{y})$ is the
unitary operator for the evolution and $\theta$ is the phase shift. A\ main
thought is to calculate the phase sensitivity for comparing the measurement
results by taking the TF state and the state in Eq. (\ref{ntf}) as the initial
states. This gives whether the influence of gravitational waves on the TF
state can be observed. Using the error propagation formula, $(\Delta
\theta)^{2}=\frac{((\Delta J_{z}^{2})_{out})^{2}}{|d\langle J_{z}^{2}%
\rangle_{out}/d\theta|^{2}}$, the optimal phase sensitivity is given as
\cite{mwn11}
\begin{equation}
(\Delta\theta)_{opt}^{2}=\frac{\Delta J_{z}^{2}\Delta J_{x}^{2}}{2(\langle
J_{x}^{2}\rangle-\langle J_{z}^{2}\rangle)^{2}}+\frac{V_{xz}}{4(\langle
J_{x}^{2}\rangle-\langle J_{z}^{2}\rangle)^{2}}, \label{ops}%
\end{equation}
where the optimal phase shift satisfies $tan^{2}\theta_{opt}=\Delta J_{z}%
^{2}/\Delta J_{x}^{2}$, $V_{xz}=\langle(J_{x}J_{z}+J_{z}J_{x})^{2}\rangle
_{i}+\langle J_{z}^{2}J_{x}^{2}+J_{x}^{2}J_{z}^{2}\rangle_{i}-2\langle
J_{z}^{2}\rangle_{i}\langle J_{x}^{2}\rangle_{i}$, and $(\Delta J_{x}%
)^{2}=\langle J_{x}^{2}\rangle-\langle J_{x}\rangle^{2}$. When the initial
state takes $\rho_{t}$ in Eq. (\ref{ntf}), the results are presented in Fig.
3. From the Fig. 3, it is seen that the peak of the phase sensitivity curves
will be shifted toward larger frequency when the amplitude of gravitational
waves decreases, which has similar behaviors to that for spin-squeezing
parameters in Fig. 1. Moreover, the increase of atomic number can improve the
phase sensitivity as seen in Fig. 4 as expected.

When the initial state takes TF state without being influenced by
gravitational waves, the phase sensitivity is obtained as
\begin{equation}
(\Delta\theta)_{TF}^{2}=1/(2j(j+1)),
\end{equation}
which gives the phase sensitivity with $\sqrt{\frac{2}{N(N+2)}}$ approaching
the the Heisenberg limit \cite{ps09}. For $N=50$ as taken in Fig. 4, the phase
sensitivity is estimated as $(\Delta\theta)_{TF}^{2}\sim7.7\times10^{-4}$.
There would be an increase in the value of phase sensitivity after the
influence of gravitational waves. This means that it is possible to detect the
influence in experiment if the initial state is exactly prepared as the TF state.

\begin{figure}[ptb]
\centering
\includegraphics[width=1\columnwidth]{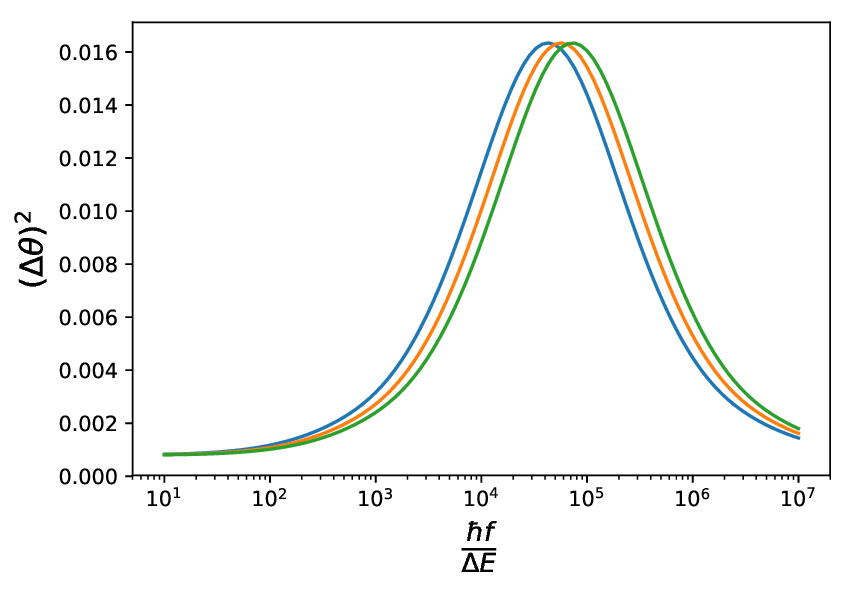}\caption{A plot of the phase
sensitivity of quantum multibody states influenced by gravitational waves for
different amplitude $h$. Blue, orange and green lines stand for $h = 10^{-18}%
$, $h = 10^{-21}$, and $h = 10^{-24}$, respectively. Other parameters are the
same as Fig. 1.}%
\label{Fig3}%
\end{figure}

\begin{figure}[ptb]
\centering
\includegraphics[width=1\columnwidth]{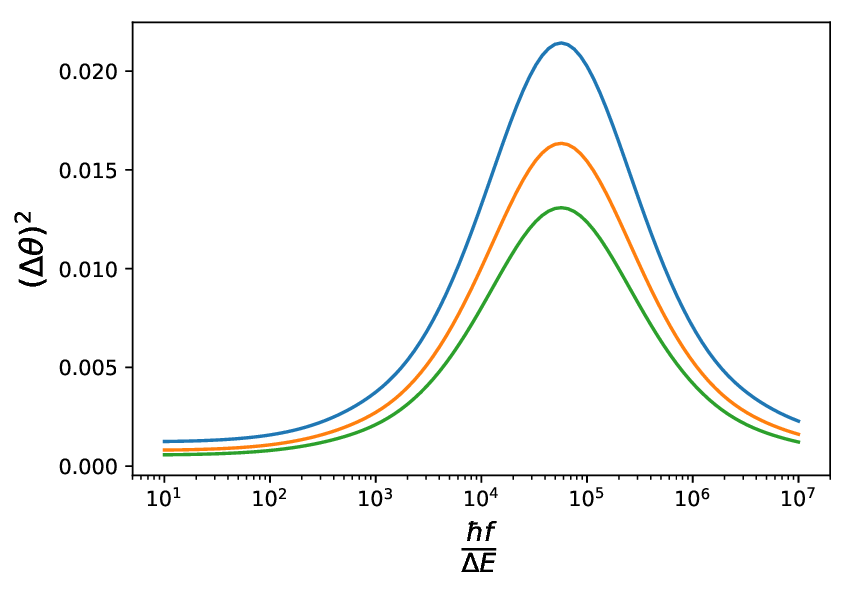}\caption{A plot of the phase
sensitivity of quantum multibody states influenced by gravitational waves as a
function of the frequency for different atom number. Blue, orange and green
lines stand for $N = 40$, $N = 50$, and $N = 60$, respectively. Other
parameters are the same as Fig. 1.}%
\label{Fig4}%
\end{figure}

But actually, a standard TF state is hard to be prepared in the present
experimental conditions. In the experiment in \cite{lty17}, there are actually
some factors leading to a deviation of standard TF state, e.g. the influence
of nonaidabatic excitations and the atom loss which causes the quantum
multibody state to be a mixture of Dicke state but not a standard TF state.
For simplicity, we consider the mixture of Dicke state as $\rho_{i}=\sum
_{j,m}p_{j,m}|j,m\rangle\langle j,m|$ where $p_{j,m}=p_{j}p_{m}$, and both
$p_{j}$ and $p_{m}$ satisfy the Gaussian distribution, $p_{j}=\frac{1}%
{\sqrt{2\pi}\sigma_{j}}e^{-\frac{(j-\overline{j})^{2}}{2\sigma_{j}^{2}}}$,
$p_{m}=\frac{1}{\sqrt{2\pi}\sigma_{m}}e^{-\frac{m^{2}}{2\sigma_{m}^{2}}}$
($\sigma_{j},\sigma_{m}\ll\overline{j}$). When all atoms are influenced by the
same gravitational waves, the Dicke state $|a,b\rangle$ becomes
\begin{equation}
\rho_{Dicke}=\sum_{m=-a}^{a}B_{m}^{2}|a,m\rangle\langle a,m|,
\end{equation}
where $B_{m}^{2}=\sum_{k=0}^{\min{a-m,a+b}}C_{a-b}^{m-b+k}C_{a+b}^{k}%
P_{+}^{m-b+k}P_{-}^{k}$ for $m\geq b$ and $B_{m}^{2}=\sum_{k=0}^{\min
{a+m,a-b}}C_{a-b}^{k}C_{a+b}^{b-m+k}P_{+}^{k}P_{-}^{b-m+k}$ for $m<b$.

We have calculated the change of entanglement related spin-squeezing parameter
for a mixture of Dicke state influenced by gravitational waves. It is clear
from Fig. 5 that for both of the two initial states, the spin-squeezing
parameter first increases and then decreases with growing frequency, i.e., the
amount of quantum entanglement for quantum-multibody states first decreases to
a specific value and then increases slowly. Given that the spin-squeezing
parameter for an initial mixed Dicke state is not zero at first, the change of
spin squeezing for a mixed Dicke state seems less than a TF state, which means
that it is more difficult to detect gravitational waves using a mixture of
Dicke states, so it is important to prepare standard TF state as gravitational
waves detectors.

When the real experimental values are used, the changes of spin squeezing
induced by gravitational waves are so small that we actually cannot discern
that with a TF state or mixed Dicke state of $11,000$ atoms, but if the atomic
number increases, the influence of gravitational waves can be observed
obviously by the change of the phase sensitivity. For example, the atomic
number is given as $10^{28}$ as in the recent suggestion using BEC to detect
gravitational waves \cite{ram19}, the phase sensitivity will increase $5$
orders of magnitude before and after the influence of gravitational waves for
the initial TF state according to our calculation in Eq. (\ref{ops}). There is
another method to improve the sensitivity by decreasing the spin-squeezing
parameter or increasing entanglement of the initial multibody quantum states.
This is shown in Fig. 6, which gives the sensitivity curves for detecting
gravitational waves with different spin-squeezing parameters. It shows that
the initial quantum state should be close enough to the TF state, which is
still out of the present experimental range.

\begin{figure}[ptb]
\centering
\includegraphics[width=1\columnwidth]{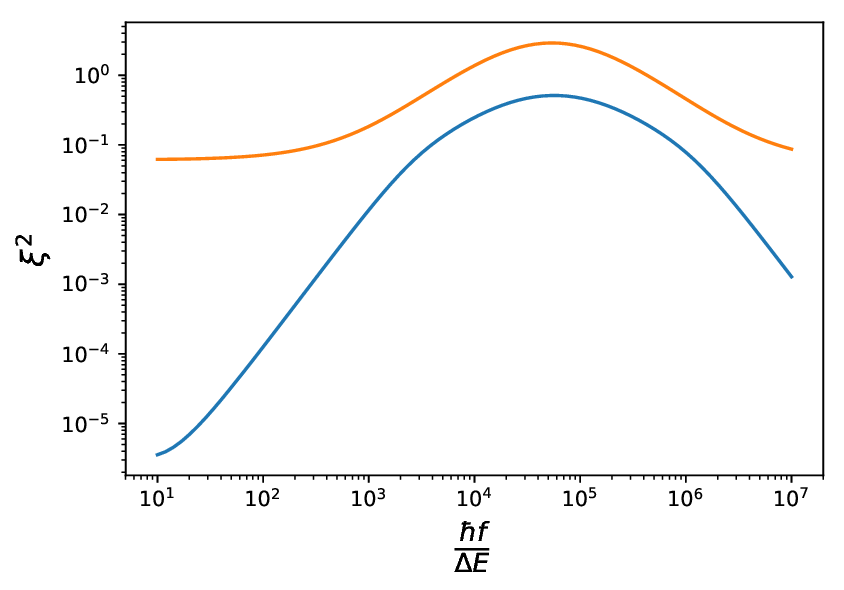}\caption{A plot of the spin
squeezing parameter of quantum multi-body states influenced by gravitational
waves as a function of the frequency. Blue and orange lines stand for TF state
and mixed Dicke state respectively. We set atom number $N = 20,\overline{j} =
10, \overline{m} = 0, \sigma_{j} = 0.5, \sigma_{m} = 0.5$, and other
parameters are the same as Fig.1.}%
\label{Fig5}%
\end{figure}

\begin{figure}[ptb]
\centering
\includegraphics[width=1\columnwidth]{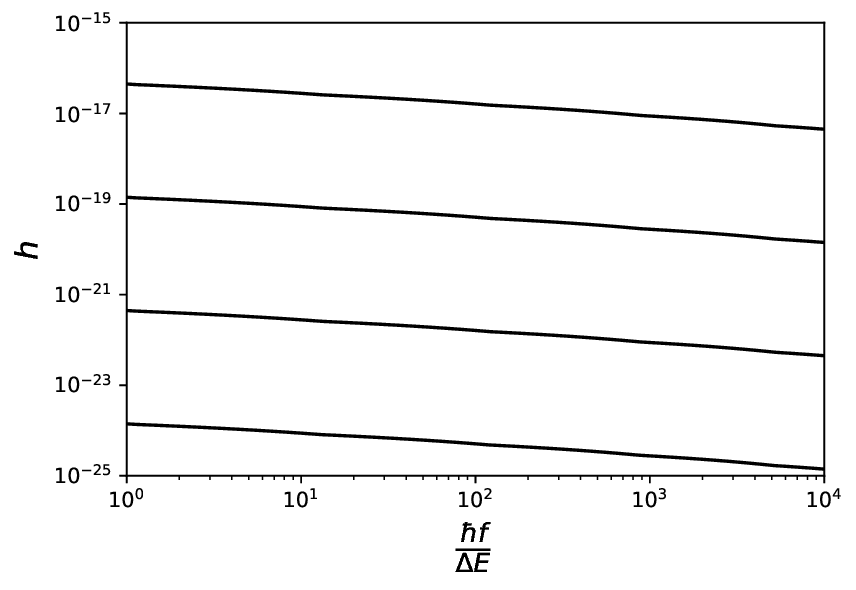}\caption{A plot of the spin
squeezing parameter of TF states influenced by gravitational waves as a
function of the frequency and amplitude. We set $t_{obs} = 10^{6}\ s$, $N =
11,000$, $P_{v} = 100$ and $\Delta E = 2\pi\hbar\ J$. The four black lines
represent $\xi^{2} = 10^{-47}$, $\xi^{2} = 10^{-49}$, $\xi^{2} = 10^{-51}$,
$\xi^{2} = 10^{-53}$ respectively from top to bottom.}%
\label{Fig6}%
\end{figure}

\section{Conclusion}

In this paper, we investigate the effect of gravitational waves on a single
Unruh-Dewitt detector. We extended this work to quantum multibody states. We
study the change of entanglement for the initial TF state by the influence of
gravitational waves using the spin-squeezing parameter. It is found that there
is a peak for every curve of the spin-squeezing parameters with different
amplitudes of gravitational waves. In particular, we find that when the proper
frequency of gravitational waves is chosen, the spin-squeezing parameter
decreases with the increasing amplitude of gravitational waves. This means
that entanglement increases when the field of gravitational waves becomes
stronger and stronger, which is similar to the change of entanglement in the
anti-Unruh effect caused by the acceleration. We also estimate the feasibility
of experimental observation using the phase sensitivity, which is far from the
present conditions. Such study is worthwhile and novel for understanding the
behaviors of gravitational waves using their interaction with the quantum
physical states.

\section{Acknowledgments}

This work is supported by National Natural Science Foundation of China (NSFC)
with Grant No. 12375057, and the Fundamental Research Funds for the Central
Universities, China University of Geosciences (Wuhan) with No. G1323523064.


\begin{thebibliography}{99}                                                                                               %


\bibitem {pt08}A. Peres and D. R. Terno, Rev. Mod. Phys. \textbf{76}, 93 (2008).

\bibitem {chm08}L. C. B. Crispino, A. Higuchi, and G. E. A. Matsas, Rev. Mod.
Phys. \textbf{80}, 787 (2008).

\bibitem {mmc17}Y. Ma, H. Miao, B. H. Pang, M. Evans, C. Zhao, J. Harms, S.
Roman, Y. Chen, Nat. Phys. \textbf{13}, 776 (2017).

\bibitem {sss20}J. S\"{u}dbeck, S. Steinlechner, M. Korobko, and R.
Schnabel,\ Nat. Photonics \textbf{14}, 240 (2020).

\bibitem {edk18}E. Knyazev, S. Danilishin, S. Hild, and F. Y. Khalili, Phys.
Lett. A \textbf{382}, 2219 (2018).

\bibitem {gkt21}F. Gray, D. Kubiznak, T. May, S. Timmerman, and E. Tjoa, J.
High Energy Phys. \textbf{11} (2021) 054.

\bibitem {pdd22}J. Paczos, K. Dbski, P. T. Grochowski, A. R. H. Smith, and A.
Dragan, arXiv:2204.10609.

\bibitem {xsa20}Q. Xu, S. AliAhmad, and A. R. H. Smith, Phys. Rev. D
\textbf{102}, 065019 (2020).

\bibitem {tp22}T. Prokopec, arXiv:2206.10136.

\bibitem {hp22}R. V. Haasteren and T. Prokopec, arXiv:2204.12930.

\bibitem {yzy22}Y. Li, B. Zhang, and L. You, New J. Phys. \textbf{24}, 093034 (2022).

\bibitem {sbf14}C. Sab\'{\i}n, D. E. Bruschi, M. Ahmadi, and I. Fuentes, New
J. Phys. \textbf{16}, 085003 (2014).

\bibitem {ram19}M. P. G. Robbins, N. Afshordi, and R. B. Mann, J. Cosmol.
Astropart. Phys. \textbf{07}, 032 (2019).

\bibitem {ram22}M. P. G. Robbins, N. Afshordi, A. O. Jamison, and R. B. Mann,
Classical Quantum Gravity \textbf{39}, 175009 (2022).

\bibitem {wgu76}W. G. Unruh, Phys. Rev. D \textbf{14}, 870 (1976).

\bibitem {udm}B. S. DeWitt, S. Hawking and W. Israel, \textit{General
Relativity: an Einstein Centenary Survey}, (Cambridge University Press,
Cambridge, England, 1979).

\bibitem {mtw73}C. W. Misner, K. S. Thorne, and J. A. Wheeler,
\textit{Gravitation}, (W. H. Freeman and Company, USA, 1973).

\bibitem {bd84}N. D. Birrell and P. C. W. Davies, \textit{Quantum Fields in
Curved Space} (Cambridge University Press, Cambridge, England, 1984).

\bibitem {st86}S. Takagi, Prog. Theor. Phys. Suppl. \textbf{88}, 1 (1986).

\bibitem {rhd54}R. H. Dicke, Phys. Rev. \textbf{93}, 99 (1954).

\bibitem {mwn11}J. Ma, X. Wang, C. P. Sun, and F. Nori, Phys. Rep.
\textbf{509}, 89 (2011).

\bibitem {sdz01}A. S\o rensen, L. Duan, J. Cirac, and P. Zoller, Nature
(London) \textbf{409}, 63 (2001).

\bibitem {tkp09}G. T\'{o}th, C. Knapp, O. G\"{u}hne, and H. J. Briegel, Phys.
Rev. A \textbf{79}, 042334 (2009).

\bibitem {gt09}O. G\"{u}hne and G. T\'{o}th, Phys. Rep. \textbf{474}, 1 (2009).

\bibitem {bmm16}W. G. Brenna, R. B. Mann, and E. Mart\'{\i}n-Mart\'{\i}nez,
Phys. Lett. B \textbf{757}, 307 (2016).

\bibitem {lzy18}T. Li, B. Zhang, and L. You, Phys. Rev. D \textbf{97}, 045005 (2018).

\bibitem {bgu17}A. Bassi, A. Gro\ss ardt, and H. Ulbricht, Classical Quantum
Gravity \textbf{34}, 193002 (2017).

\bibitem {nr85}N. Ramsey, \textit{Molecular Beams} (Oxford University Press,
Oxford, England, 1985).

\bibitem {ymk86}B. Yurke, S. L. McCall, and J. R. Klauder, Phys. Rev. A
\textbf{33}, 4033 (1986).

\bibitem {ps09}L. Pezz\'{e} and A. Smerzi, Phys. Rev. Lett. \textbf{102},
100401 (2009).

\bibitem {lty17}X.-Y. Luo, Y.-Q. Zou, L.-N. Wu, Q. Liu, M.-F. Han, M. K. Tey,
and L. You, Science \textbf{355}, 620 (2017).
\end{thebibliography}
\end{document}